\documentclass[aps,prb,twocolumn,showpacs,superscriptaddress]{revtex4}

\usepackage{graphicx}

\begin{document}

\preprint{12}

\title{First Study of Intersubband Absorption in Electrons on Helium under Quantizing Magnetic Fields}

\author{Denis Konstantinov}
\email[E-mail: ]{denis@oist.jp}
\affiliation{Okinawa Institute of Science and Technology, Tancha 1919-1, Okinawa 904-0412, Japan}
\affiliation{Low Temperature Physics Laboratory, RIKEN, Hirosawa 2-1, Wako 351-0198, Japan}
\author{Kimitoshi Kono}
\affiliation{Low Temperature Physics Laboratory, RIKEN, Hirosawa 2-1, Wako 351-0198, Japan}

\date{\today}

\begin{abstract}
We present the first measurements of inter-subband absorption of microwaves in surface electrons on liquid helium subjected to perpendicular magnetic field $B$. In quantizing $B$, the power absorption shows intermittent regions of enhanced and suppressed absorption. This behavior is caused by strong variation of the excited-electron decay rate with $B$. Particularly, fast decay due to elastic scattering provides condition for strong absorption and overheating of the electron system, while slow decay due to inelastic scattering limits absorption and causes its saturation. An unexpected feature is the strong suppression of absorption at magnetic fields where the inter-subband energy splitting is a multiple number of the cyclotron energy.  
\end{abstract}


\maketitle

Surface state electrons (SSEs) on liquid helium provide a unique example of a two-dimensional electron system~\cite{Monarkha_book}. Below 1 K, they are suspended in a vacuum about 10~nm above the liquid because of a strong repulsive barrier at the free surface. An attractive image potential is responsible for a subband structure with a discrete 1D energy spectrum $\epsilon_n=R/n^2$, where $n=1,2,..$ and $R=7.6$ (4.2)~K for liquid $^4$He ($^3$He). Such electrons are strongly-correlated~\cite{Ikegami2009,Rousseau2009,Konstantinov2009,Rees2012}, very mobile,~\cite{Bradbury2011,Rees2011} and interact very weakly with the environment mostly through coupling to surface capillary waves (ripplons). Great interest in SSEs comes from several proposals to use electrons interacting with resonant microwave radiation for quantum computing as this system promises a scalable architecture for qubits with a very long coherence time~\cite{Platzman1999,Lea2000,Dahm2002,Lyon2006,Mostame2008,Shuster2010,Wei2010}. In particular, these proposals stimulated recent studies of the inter-subband microwave absorption in SSEs~\cite{Collin2002,Isshiki2007,Konstantinov2007}. Collin et al. found that temperature-dependent contribution to the absorption linewidth $\gamma$ is small at low $T$ and obtained a relatively large value of $\Omega_R/\gamma\approx 300$ ($\Omega_R$ is the Rabi frequency), which estimates the number of qubit operations per coherence time~\cite{Collin2002}. The Rabi frequency was estimated from the observed saturation of absorption assuming that electrons populate only the two lowest surface states. Later, it was shown that the excited-electron decay due to elastic scattering from the vapor atoms or ripplons provide an efficient channel to transfer the excitation energy into the kinetic energy of electron in-plane motion~\cite{Konstantinov2007}. This causes strong overheating of the electron system, which results in population of higher excited states and absorption bleaching. The latter produces features similar to the absorption saturation in a two-level system but can be adequately described only taking into account the contribution from many energy levels~\cite{Konstantinov2007}. 
\newline
\indent The situation is different in a quantizing magnetic field $B$ applied perpendicular to the surface, which confines the in-plane states of electrons to Landau levels $E_l=\hbar\omega_c(l-1/2)$, $l=1,2,..$, where $\omega_c=eB/m$ is the cyclotron frequency. Then, the fast decay of the microwave-excited electron due to elastic scattering can be suppressed providing that $\epsilon_2-\epsilon_1\neq \hbar\omega_c l$ and broadening of Landau levels do not exceed $\hbar\omega_c$. Similar suppression of the decay rate was also predicted for localized qubits with quantized energy of in-plane motion and is a necessary condition for long coherence of qubits~\cite{Platzman1999}. When elastic scattering is suppressed, the decay rate is limited by slow inelastic processes, such as an emission of two short-wavelength ripplons, and the absorption rate must quickly saturate due to distribution of electrons between the two lowest surface states with nearly equal occupancies. Correspondingly, in experiments one expects to see strong suppression of the absorption at magnetic fields far from commensurability values $B_l$ such that $(\epsilon_2-\epsilon_1)/\hbar\omega_c\approx \omega/\omega_c=l$, where $\omega$ is the angular frequency of resonant microwaves. To the best of our knowledge, the experimental observation of this predicted behavior has not been reported yet. 
\newline  
\indent More interest in absorption experiments under quantizing $B$ comes from some remarkable microwave-induced transport phenomena, such as zero-resistance states (ZRS), recently found in electrons on helium~\cite{Konstantinov2010} as well as in two-dimensional electron gas in GaAs/AlGaAs heterostructures~\cite{Mani2002,Zudov2003,Yang2003}. In these experiments the longitudinal conductivity ($\sigma_{xx}$) and resistivity ($\rho_{xx}$) of electrons under irradiation oscillate periodically in $\omega/\omega_c\propto B^{-1}$ ($\omega$ is the angular frequency of microwaves) and vanish at the oscillation minima. Transport phenomena have been extensively studied in both systems using various techniques~\cite{Dorozhkin2011,Andreev2011,KonstantinovJPSJ2012}. However, except for a preliminary measurement of microwave transmission through a waveguide containing GaAs/AlGaAs samples~\cite{Mani2004}, no study of microwave absorption has been reported in conjunction with photo-induced conduction oscillations and ZRS.
\newline
\indent Here, we present the first measurements of $n=1\rightarrow 2$ intersubband absorption of microwaves in electrons on helium in quantizing magnetic fields. The oscillations of the absorption follow the variation of $B$-dependent elastic scattering of electrons between the two subbands. The absorption shows a complicated dependence on the magnetic field that strongly deviates from the corresponding dependence of $\sigma_{xx}$. Increased absorption near integral values of $\omega/\omega_c$ indicates overheating of electrons. The electron temperature is estimated by taking into account the temperature dependence of a many-electron fluctuating electric field. It is shown that strong suppression of absorption far from integral values of $\omega/\omega_c$ arises from the absorption saturation. To the best of our knowledge, this is the first observation of absorption saturation in this system.
\newline
\begin{figure}[t]
\centering
\includegraphics[width=8.5cm]{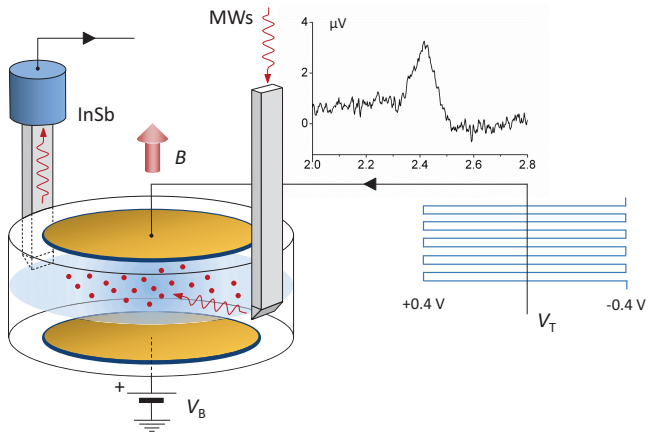}
\caption{\label{fig:1}(color online). Schematic diagram of the experimental method. Electrons on helium are contained between two circular electrodes and driven on- and off-resonance with applied microwaves by alternating the voltage at the top electrode $V_{T}$ between $\pm0.4$~V. The absorption signal is detected with a InSb Putley bolometer. In the inset: The bolometer signal versus $V_B-0.4$~V (conversion factor to frequency $\sim 3$~GHz/V).} 
\end{figure}
\indent The electron system is formed on the surface of liquid $^3$He contained in a flat cylindrical cavity as schematically shown in Fig.~\ref{fig:1}. Microwaves of fixed frequency ($\omega/2\pi=78.7$~GHz in this experiment) are introduced into the cavity through a waveguide and exit through an output port coupled to a low-temperature InSb bolometer. Intersubband transitions of electrons are resonantly excited by tuning the transition frequency $\omega_{21}=(\epsilon_2-\epsilon_1)/\hbar$ through the linear Stark shift in a perpendicular electric field $E_{\perp}$. The latter is formed by applying a constant positive potential $V_{B}$ to a circular electrode of 20 mm diameter placed underneath and parallel to the liquid surface. A similar electrode is placed above the liquid and kept at zero dc potential. The distance between electrodes is about 2.6 mm. In order to measure the absorption of microwaves by SSEs, a square wave (1~kHz) potential is applied to the top electrode to alternatively drive SSEs on- and off-resonance. Therefore, each half-cycle the microwave power reaching the bolometer is decreased by an amount absorbed by SSEs, and the bolometer signal proportional to power absorption is measured as a function of a perpendicular magnetic field using the conventional lock-in technique. 
\newline
\begin{figure}[t]
\centering
\includegraphics[width=8.5cm]{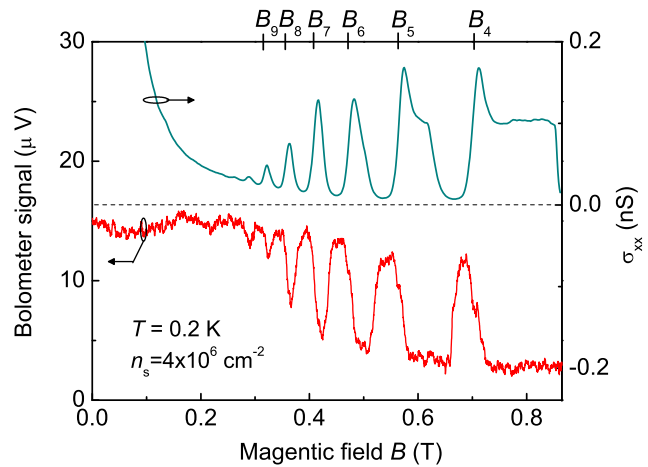}
\caption{\label{fig:2}(color online). An amplified ($\times 500$) bolometer signal (lower curve, red) and longitudinal conductivity $\sigma_{xx}$ (upper curve, green) versus magnetic field $B$ for $T=0.2$~K and $n_s=4.0\times 10^6$~cm$^{-2}$. The top horizontal axis shows the values of $B$, such that $\omega/\omega_c=l$, for $l=4$, 5, 6, 7, 8 and 9.} 
\end{figure}
\indent The bolometer signal measured at the temperature of the liquid $T=0.2$~K is shown in Fig.~\ref{fig:2}. Within the random noise and slow drift of the bolometer signal, absorption remains constant for $B\lesssim 0.2$~T. Above this field, absorption starts oscillating as a function of $B$. The onset field coincides with that for conductance oscillations reported previously~\cite{Konstantinov2010}. An example of $\sigma_{xx}$ versus $B$ measured at the same $T$, while electrons are continuously tuned to the resonance with microwaves, is also shown in Fig.~\ref{fig:2}. Intuitively, the variation of both absorption and transport properties as a function of $B$ must result from the Landau quantization and the peak structure of the density-of-state function for each subband. In particular, the intersubband scattering of electrons, which is predominantly elastic in zero magnetic field, is significantly suppressed in magnetic fields away from values of $B_l$ satisfying the commensurability condition $\epsilon_2 - \epsilon_1 = \hbar\omega_c l$, that is, when the density-of-state functions of the two subbands do not overlap~\cite{Konstantinov2010}. As shown below, this strongly affects the power absorption at resonance. Note that the shape of conductance oscillations significantly differs from that of the power absorption. In the vicinity of $B_l$, the conductivity has maximum (minimum) at positive (negative) $(B-B_l)$. According to the linear transport theory of irradiated electrons recently proposed by Monarkha~\cite{Monarkha2011}, this behavior of $\sigma_{xx}$ is attributed to the appearance of an "anomalous" contribution to the effective collision rate of electrons due to nonequilibrium occupancy of the second subband. The anomalous contribution changes sign as $B$ passes through $B_l$ and produces the deep minima of $\sigma_{xx}$ appearing to the left of $B=B_l$. On the other hand, the power absorption shows strong enhancement in the vicinity of $B_l$ for both positive and negative $(B-B_l)$. 
\newline
\indent Note that the enhancement in absorption should lead to electron overheating, as would follow from the balance between the energies absorbed by electrons from the radiation field and lost in the inelastic collisions with scatterers~\cite{Konstantinov2007}. The exact calculation of the energy balance in quantizing magnetic fields is difficult, owing to the complicated dynamics of strongly interacting electrons, as will be discussed later. Fortunately, it is possible to roughly estimate electron overheating by considering the effect of a many-electron fluctuating electric field $\textbf{E}_f$ acting on an electron~\cite{Dykman1997}. This field appears as a result of the Coulomb interaction between electrons and has a probability distribution with a Gaussian shape of width $\langle E_f \rangle \propto \sqrt{T_e} n_s^{3/4}$, where $T_e$ is the electron temperature. The energy of an electron scattered by a distance $(X-X')$ ($X$ is the in-plane coordinate of the cyclotron orbit center) changes by $e |\textbf{E}_f| (X-X')$.  Consequently, the maxima of the density-of-states functions for each subband must be shifted by the same amount to realize maximum scattering. In particular, the periodic structure of the density-of-states due to the Landau quantization has little effect on the average scattering rate unless $\hbar\omega_c$ exceeds $e\langle E_f \rangle (X-X')$. For classical orbits, the scattering distance is approximately the cyclotron radius $R_B=\sqrt{2mk_BT_e}/eB$, which determines the onset field $B_0\propto \sqrt{T_e}n_s^{3/8}$~\cite{Dykman1997}. Taking the experimental conditions in Fig.~\ref{fig:1}(a) and assuming $T_e\approx T$, this field is of order 0.1~T. The actual onset field of about 0.25~T that can be seen in Fig.~\ref{fig:2} indicates the overheating of the electrons to the temperature $T_e\sim 1$~K at $B<B_0$, as well as in the vicinity of the commensurability fields $B_l$, $l=4$, 5, 6, 7, 8 and 9, where absorption shows strong enhancements. An increase in $n_s$ has a similar effect to an increase in $T_e$ and provides an effective means of verifying this model. The oscillations for surface densities of $n_s=3.5\times 10^6$ and $2.0\times 10^6$~cm$^{-2}$ are shown in Fig.~\ref{fig:3} where the bolometer signal is plotted versus $\omega/\omega_c$. We expect that the onset of oscillations versus $\omega/\omega_c$ increases as $n_s^{-3/8}$. The ratio of this quantity is about 1.2 for the two densities in Fig.~\ref{fig:3}. This is in good agreement with the ratio of the onset fields for the two traces shown in this figure.
\newline
\begin{figure}[t]
\centering
\includegraphics[width=8.5cm]{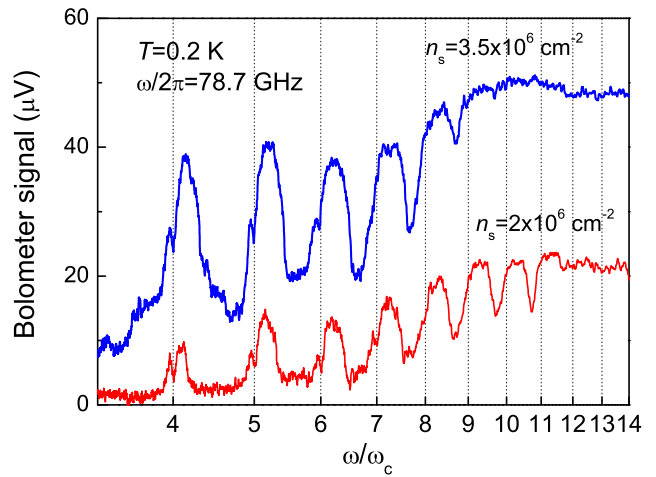}
\caption{\label{fig:3}(color online). An amplified ($\times 10^3$) bolometer signal versus the ratio $\omega/\omega_c$ for $T=0.2$~K and two values of surface density $n_s=3.5\times 10^6$ (top curve) and $2.0\times 10^6$ cm$^{-2}$ (bottom curve).} 
\end{figure}
\indent As seen in Fig.~\ref{fig:3}, the absorption signal is strongly suppressed when far from integral values of $\omega/\omega_c$. The reason for such a remarkable decrease of absorption is its saturation when the rate of non-radiative decay of excited-state electrons is much smaller than the rate of their induced emission. The average rate of photon absorption per electron $\alpha$ at resonance is given by the difference of the absorption and induced emission rates, that is $\alpha = (\Omega_R^2/\gamma)(\rho_1-\rho_2)$, where $\Omega_R$ is the Rabi frequency determined by the power of the radiation field and $\gamma$ is the width of the resonance. The fractional occupancies of subbands $\rho_n$ are determined by the balance of the scattering of electrons between different subbands. If we neglect the thermal population of higher subbands, which is valid for the case of moderate overheating estimated above, the occupancies of the two lowest subbands satisfy $\rho_1+\rho_2=1$ and $\rho_1 \Omega_R^2/\gamma \approx \rho_2 (\Omega_R^2/\gamma+\nu_{21})$, where $\nu_{21}$ is the rate of excited-electron decay. From this, we obtain $\alpha\approx \nu_{21}\Omega_R^2/(2\Omega_R^2+\nu_{21}\gamma)$. The absorption rate strongly depends on the relation between $\Omega_R$, $\gamma$ and $\nu_{21}$. In the described experiment, the value of $\Omega_R$ is determined by the input power of microwaves and is fixed. The typical value of $\Omega_R$ was estimated earlier and is on the order of $10^8$~s$^{-1}$~\cite{Konstantinov2007}. As seen in the inset of Fig.~\ref{fig:1}, the width of the resonance $\gamma\approx 2\times 10^9$~s$^{-1}$. Most probably it is determined by inhomogeneous broadening due to, for example, spatial variation of the perpendicular electric field $E_{\perp}$ across the electron system~\cite{Isshiki2007}. Thus, it is also fixed in the experiment. On the other hand, the decay rate $\nu_{21}$ varies strongly with $B$. Near commensurability fields, $B_l$, the decay is predominantly due to elastic scattering from a single ripplon. At $T=0.2$~K, the decay rate $\nu_{21}\approx 10^8$~s$^{-1}$~\cite{Monarkha2012}. Thus, the average absorption rate $\alpha\approx \Omega^2/\gamma\sim 10^7$~s$^{-1}$. Note that under such conditions $\rho_2\ll \rho_1 \approx 1$. Far from commensurability fields, elastic scattering is strongly suppressed and the decay rate is limited by the inelastic process of two-ripplon emission~\cite{Monarkha2010}. Using the theoretically predicted value of $\nu_{21}\approx 10^6$~s$^{-1}$ of the inelastic scattering rate~\cite{Monarkha2010}, we have $\nu_{21}\ll \Omega_R$. Thus we conclude that $\alpha\approx \nu_{21}/2\sim 10^6$~s$^{-1}$, which is independent of $\Omega_R$. This corresponds to the absorption saturation with nearly equal occupancies of the two lowest subbands, $\rho_2\rightarrow \rho_1\approx 1/2$. Note that the absorption rate is about order of magnitude smaller than $10^7$~s$^{-1}$ estimated for the absorption rate near commensurability fields. This is in agreement with the experimental result. 
\newline
\indent An interesting feature of our results is the appearance of narrow dips in power absorption centered at the integral values of $\omega/\omega_c$, as seen in Fig.~\ref{fig:3}. The origin of this effect could be related to a possible variation of the intersubband transition rate $\Omega_R^2/\gamma$. For example, if the dominant contribution to the absorption width $\gamma$ comes from homogeneous broadening, it varies strongly with the magnetic field. In general, homogeneous broadening contains contributions from scattering of electrons within subbands, as well as excited-electron decay~\cite{Ando1978}. When $\gamma\approx \nu_{21}$, from the above equations we obtain $\alpha\approx \nu_{21}\Omega_{21}/(2\Omega_{21}^2+\nu_{21}^2)$. It is easy to see that if $\nu_{21}$ exceeds $\sqrt{2}\Omega_R$, the absorption rate would decrease and would attain minima when $\nu_{21}$ reaches maximal values at $B=B_l$. This effect was theoretically predicted by Monarkha, who considered the variation of electron temperature with magnetic field by solving the energy balance equation (c.f. Figs. 1 and 2 from Ref.~26). However, in our experiments inhomogeneous broadening is significantly larger than the expected homogeneous contribution. Because it is $B$-independent broadening, the absorption rate $\alpha$ is a monotonically increasing function of $\nu_{21}$. Therefore, it seems unlikely that the above mechanism is responsible for the dip structure seen in Fig.~\ref{fig:3}. Also, the theory of Ref.~26 predicts stronger overheating of SSEs on the high-field side of the commensurability fields $B_l$ compared with the low-field side, due to anomalous contribution to energy dissipation. This results in asymmetric peaks (with respect to $B=B_l$) of $T_e$ and, therefore, power absorption. However, the observed peaks of power absorption have the opposite asymmetry (see Figs.~\ref{fig:2} and \ref{fig:3}) with significantly stronger absorption on the low-field side of $B_l$. While we cannot clarify the origin of this strong asymmetry at the moment, we would like to point out a similar asymmetry in the amplitudes of maxima and minima of $\sigma_{xx}$ in the vicinity of $B_l$.  
\newline
\indent It is important to note that the observed broadening of the absorption line is at least one order of magnitude larger than the homogeneous linewidth. This means that in the experiment most of the electrons (at least 90$\%$) are detuned from resonance with the applied microwaves. Such a large detuning should have strong effect on the anomalous contribution to longitudinal conductivity because unbalanced negative momentum transfered by decaying electrons appears only when the excited-subband occupancy differs from thermal occupancy~\cite{Monarkha2011}. Thus, the effect of negative conductivity must be localized within the regions of the applied electric field $E_{\perp}$ the magnitude of which satisfies conditions for the inter-subband resonance. Recent experiments with SSEs on helium showed that the effect of ZRS is accompanied by a spatial redistribution of electrons within the whole electron system~\cite{KonstantinovJPSJ2012}. Such redistribution can be attributed to the instability of regions with $\sigma_{xx}<0$ due to growing fluctuations of electronic charge~\cite{Monarkha2012}. However, it is not immediately clear how instability can cause such a global redistribution of charge if negative conductivity regions are indeed localized.    
\newline
\begin{figure}[t]
\centering
\includegraphics[width=8.5cm]{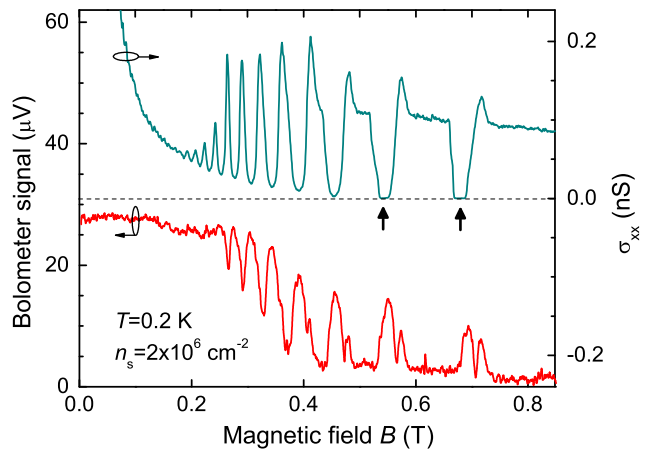}
\caption{\label{fig:4}(color online). Bolometer signal (lower curve, red) and longitudinal conductivity $\sigma_{xx}$ (upper curve, green) versus magnetic field $B$ for $T=0.2$~K and $n_s=2.0\times 10^6$~cm$^{-2}$. Zero-resistance states are indicated by bold arrows.} 
\end{figure}
\indent Interestingly, the overall behavior of power absorption described earlier does not change significantly when the irradiated SSEs exhibit ZRS. In particular, for low density data shown in Fig.~\ref{fig:4} zero resistance appears at the minima of $\sigma_{xx}$ at $B\approx 0.54$ and $0.68$~T. This points out that the spatial redistribution of charge which accompanies the formation of ZRS does not alter absorption. Again, this result is hard to understand if we assume that inhomogeneous broadening dominates because a large redistribution of electrons must significantly alter their resonant conditions. However, it might be difficult to compare the measurements of power absorption and magnetoconductivity in the ZRS regime because of the difference in characteristic time scales. On the one hand, absorption measurements were done under fast switching of resonant conditions induced by a 1~kHz-modulation of the electric field $E_{\perp}$. On the other hand, it was shown previously that the charge redistribution can build up on a time scale of several hundred milliseconds~\cite{KonstantinovJPSJ2012}. Thus, in the absorption measurements the redistribution of SSEs is expected to be smaller.   
\newline
\indent In summary, we measured absorption of microwaves due to inter-subband $n=1\rightarrow 2$ resonance in electrons on helium in quantizing magnetic fields. We conclude that the variation of power absorption with the magnetic field is mainly due to variation of the inter-subband scattering rate. The analysis of experimental data shows that the contribution of the many-electron fluctuating electric field is significant even at low densities on the order of $10^6$~cm$^{-2}$, typical for our experiments. Power absorption is large under fast decay of excited electrons due to elastic scattering. Correspondingly, the electron system is overheated to temperature $T_e\sim 1$~K. On the other hand, absorption is strongly suppressed when it is limited by a slow inelastic decay of the excited electrons. Under this condition, absorption is saturated. The estimated suppression of power absorption is consistent with the experimental results. Some features, such as strong suppression of absorption at the integral values of $\omega/\omega_c$, can not be explained by the existing theories and require further consideration.
\newline
\indent We thank Yu.~P. Monarkha for helpful discussions. The work is partially supported by Grant-in-Aids for scientific research from MEXT. D.~K. is currently supported by an internal grant from the Okinawa Institute of Science and Technology (OIST) Graduate University.

\end{document}